\documentclass[11pt]{article}

\usepackage{moreverb,url}
\usepackage{amsmath,amssymb}
\usepackage{dsfont}
\usepackage{lscape}
\usepackage{color}
\usepackage{multirow}
\usepackage{booktabs}
\usepackage{authblk}
\usepackage{graphicx}
\usepackage[text={174mm,258mm},%
papersize={210mm,297mm},%
columnsep=12pt,%
headsep=21pt,%
centering]{geometry}
\usepackage[colorlinks,bookmarksopen,bookmarksnumbered,citecolor=red,urlcolor=red]{hyperref}

\begin{document}

\title{G-computation and doubly robust standardisation for continuous-time data: a comparison with inverse probability weighting}

\date{}

\author[1,2]{Arthur Chatton\thanks{Corresponding author: \tt{arthur.chatton@univ-nantes.fr}}} 
\author[1,2]{Florent Le Borgne}
\author[3,4]{Clémence Leyrat} 
\author[,1,5]{Yohann Foucher}

\affil[1]{INSERM UMR 1246 - SPHERE, Nantes University, Tours University, Nantes, France}
\affil[2]{IDBC-A2COM, Pac\'e, France}
\affil[3]{Department of Medical Statistics, London School of Hygiene and Tropical Medicine, London, UK}
\affil[4]{Inequalities in Cancer Outcomes Network (ICON), London School of Hygiene and Tropical Medicine, London, UK}
\affil[5]{Centre Hospitalier Universitaire de Nantes, Nantes, France}

\maketitle

\begin{center}
   \LARGE{Accepted for publication in \\ \textit{Statistical Methods in Medical Research}}
\end{center}

\vspace{2cm}

\section*{Abstract}
In time-to-event settings, g-computation and doubly robust estimators are based on discrete-time data. However, many biological processes are evolving continuously over time. In this paper, we extend the g-computation and the doubly robust standardisation procedures to a continuous-time context. We compare their performance to the well-known inverse-probability-weighting (IPW) estimator for the estimation of the hazard ratio and restricted mean survival times difference, using a simulation study. Under a correct model specification, all methods are unbiased, but g-computation and the doubly robust standardisation are more efficient than inverse probability weighting. We also analyse two real-world datasets to illustrate the practical implementation of these approaches. We have updated the {\tt{R}} package {\tt{RISCA}} to facilitate the use of these methods and their dissemination.

\noindent \textbf{Keywords:} Causal inference, Parametric g-formula, Propensity score, Restricted mean survival time, Simulation study.

\section{Introduction}
Real-world evidence is scientific evidence obtained from data collected outside the context of randomised clinical trials.\cite{Sherman_RWE_2016} The absence of randomisation complicates the estimation of the marginal causal effect (hereafter referred to merely as causal effect) of exposure (including treatment or intervention) due to a potential risk of confounding.\cite{Hernan_2004} \textcolor{black}{Rosenbaum} and Rubin\cite{rosenbaum_central_1983} introduced the propensity score (PS) as a tool for causal inference in the presence of measured confounders. In a binary exposure setting, it has been shown that the estimated PS is a balancing score, meaning that conditional on the estimated PS, the distribution of covariates is similar for exposed and unexposed patients. Following this property, the PS can be used in four ways to provide estimates of the causal exposure effect: matching, stratification, adjustment, and inverse-probability-weighting (IPW).\cite{robins_marginal_2000} Stratification leads to residual confounding and adjustment relies on strong modelling assumptions.\cite{lunceford_stratification_2004, Vansteelandt_Daniel_2014} Although matching on PS has long been the most popular,\cite{ali_reporting_2015} IPW appears to be less biased and more precise in several studies.\cite{le_borgne_comparisons_2015, Hajage__2016, austin_performance_2013} Moreover, King and Nielsen\cite{King_PSM} argued for halting the use of PS matching for many reasons, including covariate imbalance, inefficiency, model dependence, and bias. \textcolor{black}{Indeed, matching on the PS is limited by the exclusion of patients without a suitable match leading to a non-representative population because of a change in the covariates distribution and a loss of statistical power.}

Causal effects can also be estimated using the g-computation (GC), a maximum likelihood substitution estimator of the g-formula.\cite{Robins_1986, Snowden_2011} While IPW is based on exposure modelling, the GC relies on the prediction of the potential outcomes for each subject under each exposure status. \textcolor{black}{Extensions of GC with time-to-event outcomes were recently proposed in a discrete-time setting.\cite{Keil_2014, Wen_2020, Breskin_2020} Only Breskin \textit{et al.}\cite{Breskin_2020} noted an extension for continuous-time setting, \textit{i.e.}, with infinitely short time intervals,\cite{Sjolander_2016} but without investigating its properties. Discrete-time models lead to estimates that depend of the length of the intervals of time, may generate interval censoring, and are often biologically implausible.\cite{Gollob_Reichardt_1987} Furthermore, the non-collapsibility of estimands due to the self-induced selection bias increases with the length of the intervals of time.\cite{Sjolander_2016}}  

With time-to-event outcomes, the presence of right-censoring and its magnitude is of prime importance since a small number of observed events due to censoring may impact the estimation of the outcome model involved in the GC. By contrast, the IPW may perform well as long as the number of exposed patients is sufficient to estimate the PS and the total sample size is sufficiently large to limit variability in the estimated weights. \textcolor{black}{To overcome potential model misspecifications, doubly robust estimators (DREs) were proposed. DREs combine both the GC and PS to obtain an unbiased estimate when at least one of the two working/nuisance models (\textit{i.e.}, a model needed to estimate the target parameter but not estimating it itself\cite{Kreif_DiazOrdaz_2019}) is well-specified.\cite{Lendle_2013, Tan_2007} Similarly to GC, current implementation of DREs focus on discrete-time data. Therefore, we present an extension as proposed by Vansteelandt and Keiding.\cite{Vansteelandt_Keiding_2011}}

Several studies (see \cite{Chatton_2020} and references therein) compared the IPW\textcolor{black}{, GC and DRE} in different contexts. They reported a lower variance for the GC than \textcolor{black}{both the IPW and DRE}. Nevertheless, to the best of our knowledge, no study has focused on time-to-event outcomes. 

In the present paper, we aimed to detail the statistical framework for using the GC in time-to-event analyses. We restricted our developments to time-invariant confounders\textcolor{black}{, and we refer the readers to Wen \textit{et al.}\cite{Wen_2020} for a recent study of GC with time-to-event outcomes and time-varying exposure. An equivalent framework for IPW can be found in Hern\'an \textit{et al.}.\cite{Hernan_2001}} We also compared the performances of the GC\textcolor{black}{, IPW and DRS}. The rest of this paper is structured as follows. In section 2, we detail the methods. Section 3 presents the design and findings of a simulation study. In section 4, we propose a practical comparison with two real-world applications related to treatment evaluations in multiple sclerosis and kidney transplantation. Finally, we discuss the results and provide practical recommendations to help analysts to choose the appropriate analysis method.

\section{Methods}

\subsection{Notations}

Let $(T_i,\delta_i,A_i,L_i)$ be the random variables associated with subject $i$ ($i=1,...,n)$. $n$ is the sample size, $T_i$ is the participating time, $\delta_i$ is the censoring indicator (0 if right-censoring and 1 otherwise), $A_i$ is a binary time-invariant exposure initiated at time $T=0$ (1 for exposed subjects and 0 otherwise), and $L_i=\{L_{1i},...,L_{pi}\}$ is the set of the $p$ measured time-invariant confounders. Let $S_a(t)$ be the survival function of group $A=a$ at time $t$, and let $\lambda_a(t)$ be the corresponding instantaneous hazard function. Suppose $D_a$ is the number of different observed times of event in group $A=a$. At time $t_j$ ($j=1,... , D_a$), the number of events $d_{ja}$ and the number of at-risk subjects $Y_{ja}$ in group $A=a$ can be defined as $d_{ja} = \sum_{i:t_i = t_j} \delta_i \mathds{1}(A_i=a)$ and $Y_{ja} = \sum_{i:t_i \geq t_j} \mathds{1}(A_i=a)$.

\subsection{Estimands}
\label{estimand}

\begin{figure}
\centering
\includegraphics[scale=0.8]{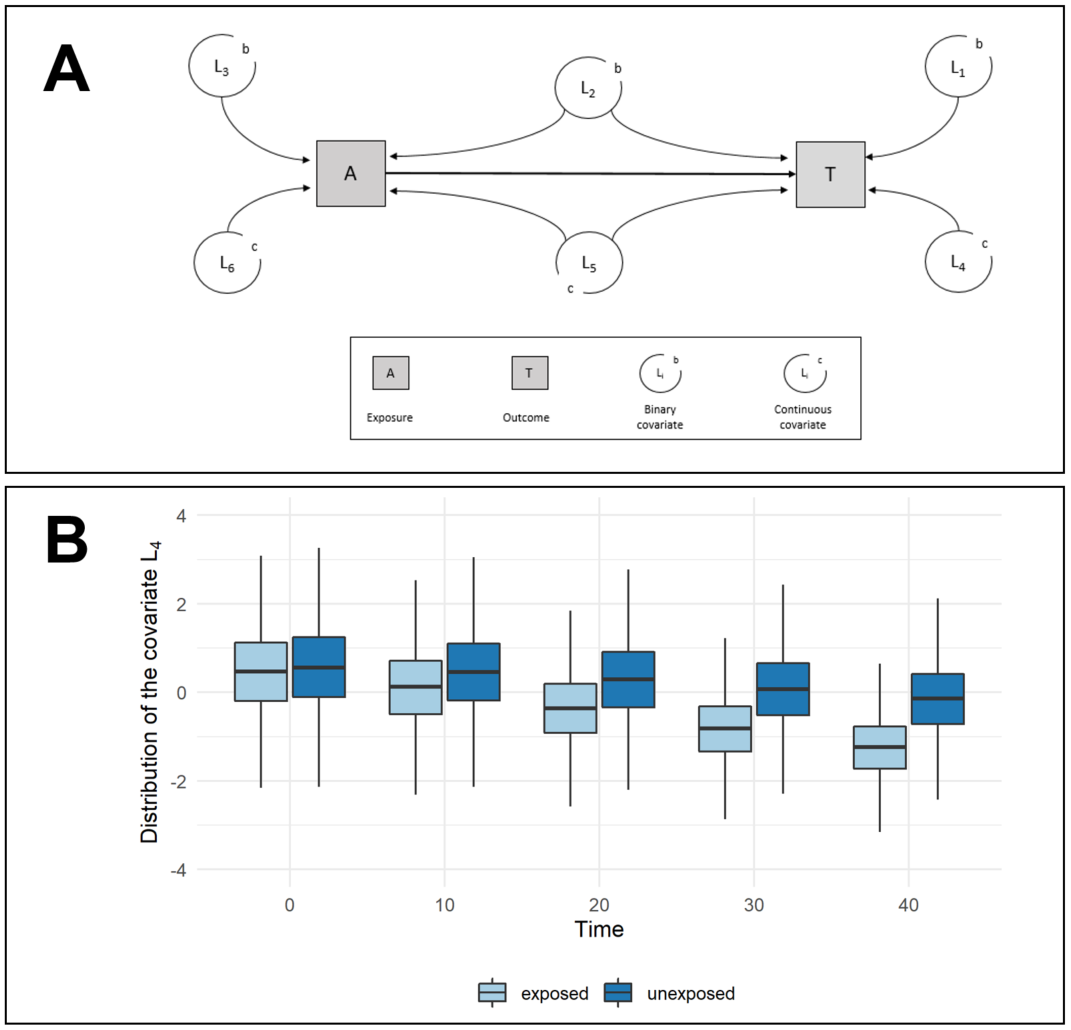}
\caption{(\textbf{A}) Causal diagram illustrating the data-generating process. (\textbf{B}) Distribution of the baseline covariate $L_4$ over time according to exposure status in a simulated population of one million people. $L_4$ is moderately associated (HR = 1.8) with the outcome.}
\label{Fig1}
\end{figure}

The hazard ratio ($HR$) has become the estimand of choice in confounder-adjusted studies with time-to-event outcomes. However, it has also been contested,\cite{hernan_hazards_2010, Aalen_2015} mainly because the \textcolor{black}{time-varying distribution of the baseline characteristics among the corresponding at-risk populations leads to selection biases. To better understand this pitfall that differs from the concept of individual time-varying covariate(s), consider the data-generating process illustrating in Figure \ref{Fig1} (panel A). If the analyst controls for confounding by adjusting or stratifying on $L_2$ and $L_5$, there is no confounding at baseline. Additionally, suppose that the individual values of the quantitative covariate $L_4$ are constant over the time, and that both the random variables $L_4$ and $A$ are independently associated with a higher risk of death. In this situation, a difference between the average $L_4$ values among the survivors appears over time (Figure \ref{Fig1}, panel B). Then, even when the conditional HR between an exposed and an unexposed with the same characteristics $L$ is constant over time, the marginal (population) HR varies over time. This is also referred to the non-collapsibility of the HR.\cite{Aalen_2015}}
Instead of $HR$, one can estimate the average over time of the different time-specific $HRs$: $AHR = \int [ \lambda_1(t) / \lambda_0(t) ] f(t)\mathrm{d}t$.\cite{Schemper_2009}

Nevertheless, Aalen \textit{et al.}\cite{Aalen_2015} concluded that it is difficult to draw causal conclusions from such a relative estimand. Hern\'an\cite{hernan_hazards_2010} advocated the use of the adjusted survival curves and related differences. For instance, the restricted mean survival time (RMST) allows us to summarise a survival curve for a specific time-window and to compare two curves by looking at the difference in RMST.\cite{Royston_Parmar_2013} The RMST difference up to time $\tau$ is formally defined as :
\begin{equation}
\Delta (\tau) = \int_{0}^{\tau} [S_1(t) - S_0(t)]\mathrm{d}t
\label{eq_rmst}
\end{equation}

This value corresponds to the difference in terms of mean event-free time between two groups of exposed and unexposed individuals followed up to time $\tau$. A further advantage of the RMST difference is its usefulness for public health decision making.\cite{Poole_2010} Note that other alternatives that might avoid this problem exist, such as the attributable fraction or the number needed to treat.\cite{Sjolander_2018}

Hereafter, we considered $AHR$ and $\Delta (\tau)$.

\subsection{Weighting on the inverse of propensity score}
\label{ss:ipw}

Formally, the PS is \textcolor{black}{defined by $g(L_i)$} $= P(A_i=1|L_i)$, \textit{i.e.}, the probability that subject $i$ is exposed according to her/his characteristics $L_i$. In practice, analysts often use a logistic regression \textcolor{black}{such that $g(L_i) = \exp(\alpha_0 + \alpha L_i)/(1+\exp(\alpha_0 + \alpha L_i))$, where $\alpha_0$ and $\alpha$ are the intercept and the regression coefficient associated with the exposure, respectively. The individual PSs are then the predictions from this model}. Let $\omega_i$ be the \textcolor{black}{stabilised} weight of subject $i$. Xu \textit{et al.}\cite{xu_use_2010} defined $\omega_i = A_i P(A_i=1) / \textcolor{black}{g(L_i)} + (1-A_i) P(A_i=0) / (1-\textcolor{black}{g(L_i)})$ to obtain a pseudo-population in which the distribution of covariates is balanced between exposure groups, enabling estimation of the causal effect in the entire population.\cite{Hernan_2004} The use of stabilised weights has been shown to produce a suitable estimate of the variance even when there are subjects with extremely large weights.\cite{robins_marginal_2000, xu_use_2010} The weighted numbers of events and at-risk subjects at time $t_j$ in group $A=a$ are $d_{ja}^\omega = \sum_{i:t_i = t_j} \omega_i \delta_i \mathds{1}(A_i=a)$ and $Y_{ja}^\omega = \sum_{i:t_i \geq t_j} \omega_i \mathds{1}(A_i=a)$, respectively. Cole and Hern\'an\cite{cole_adjusted_2004} proposed a weighted Kaplan-Meier estimator defined as:

\begin{equation}
\hat S_a(t)= \prod_{t_j \leq t}  \Big  [ 1- d_{ja}^\omega / Y_{ja}^\omega \Big ]
\label{eq1}
\end{equation}

To estimate the corresponding $AHR$, they suggested the use of a weighted univariate Cox PH model, in which exposure is the single explanatory variable. We use equation (\ref{eq_rmst}) to estimate the corresponding $\Delta(\tau)$.

\subsection{G-computation} 
\label{ss:gc}

Akin to the IPW, the GC involves two steps. The first step consists of estimating the working model \textcolor{black}{$Q(A,L)$}.\cite{Snowden_2011} When suitable, it can consist of a proportional hazard (PH) regression: $h_0(t)\exp(\gamma A_i + \beta L_i)$ where $h_0(t)$ is the baseline hazard function \textcolor{black}{at time $t$}, and $\gamma$ and $\beta$ are the regression coefficients. Estimates of the cumulative baseline hazard $\hat{H}_0(t)$ and the regression coefficients ($\hat \gamma, \hat \beta$) can be obtained by the joint likelihood approach proposed by Breslow.\cite{Breslow_1972} The second step consists of predicting the counterfactual mean survival function if all subjects would have been exposed ($do(A=1)$) or unexposed ($do(A=0)$):

\begin{equation}
\hat{S}_a(t) = n^{-1} \sum_{i=1}^n \exp \Big [- \hat{H}_0(t) \times \exp(\hat\gamma \times do(A_i=a) + \hat\beta L_i)  \Big ]
\label{eq2}
\end{equation}

Then, $\widehat{AHR}$ can be computed as the mean of the individual counterfactual hazard ratios at the observed event times:\cite{Schemper_2009}

\begin{equation}
\widehat{AHR} = \Big [ \sum_{i=1}^n \delta_i  \Big ]^{-1} \sum_{i=1}^n \delta_i \Big [\hat{\lambda}_1(t_i)/ \hat{\lambda}_0(t_i)  \Big ],
\label{eq3}
\end{equation}
where $\hat{\lambda}_a(t) = - \partial \log\hat{S}_a(t)/\partial t$, which is obtained from equation (\ref{eq2}) by numerical differentiation. We use equation (\ref{eq_rmst}) to estimate the corresponding $\Delta(\tau)$.

\subsection{\textcolor{black}{Doubly robust standardisation}}

\textcolor{black}{DREs combine $Q(A,L)$ and $g(L)$ to obtain a consistent estimate when at least one of these working models is well-specified.\cite{Tan_2007, Lendle_2013} In most cases, $g(L)$ is used to update $Q(A,L)$, such as in Targeted Maximum Likelihood Estimator (TMLE).\cite{Luque_2018} However, it can seem more intuitive to first use the IPW approach to reduce the imbalance between exposure groups and to then apply GC to control for residual confounding.\cite{Hernan_2013} Therefore, we proposed the following doubly robust standardisation (DRS).\cite{Vansteelandt_Keiding_2011} First, $g(L)$ is fitted to obtain the individual stabilised weights $\omega_i$, as defined in the subsection \ref{ss:ipw} to balance the exposure groups on $L$. Second, a weighted Q-model $Q(A,L)$ is fitted using the aforementioned weights $\omega_i$ in the procedure described in the subsection \ref{ss:gc} to achieve the double robustness property.}

\subsection{Identifiability conditions}

As for standard regression models, the IPW and the GC require assumptions of non-informative censoring, no measurement error, no model misspecification, and no interference.\cite{Hudgens_Halloran_2008} Three additional assumptions, called \emph{identifiability conditions}, are necessary for causal inference. (i) The values of exposure under comparisons correspond to well-defined interventions that, in turn, correspond to the versions of exposure in the data. (ii) The conditional probability of receiving every value of exposure depends only on the measured covariates. (iii) The conditional probability of receiving every value of exposure is greater than zero. These assumptions are known as \emph{consistency}, \emph{conditional exchangeability} and \emph{positivity}, respectively.

\section{Simulation study}

\subsection{Data generation}

We generated data in three steps \textcolor{black}{following the data-generating process illustrated in Figure \ref{Fig1} (panel A)}. (i) We simulated three covariates ($L_1$ to $L_3$) from a Bernoulli distribution with parameter equal to 0.5 and three covariates ($L_4$ to $L_6$) from a standard normal distribution. (ii) We generated the exposure $A$ according to a Bernoulli distribution with probability obtained by the logistic model with the following linear predictor: $-0.5 + \log(2) \cdot L_2 + \log(1.5) L_3 +\log(1.5) L_5 +\log(2) L_6$. We set the intercept to obtain the prevalence of exposed individuals at 50\%. (iii) We generated the times-to-event from a Weibull PH model. We set the scale and shape parameters to 40.0 and 2.0, respectively. Based on a random variable $U_i$ drawn from a standard uniform distribution, we then computed the time-to-event from a Weibull PH model as $40.0 \times [(1 - \log(1-U_i) \exp(-\gamma A_i - \log(1.3) L_1 - \log(1.8) L_2 - \log(1.8) L_4 - \log(1.3) L_5 )) - 1]^{-2.0}$, where $\gamma = \log(1.0)$ under the null hypothesis or $\log(1.3)$ under the alternative hypothesis. We subsequently censored the times-to-event using a uniform distribution on [0,70] or [0,15], leading to approximately 40\% and 90\% censored observations, respectively. For each scenario, we randomly generated 10,000 datasets.

\subsection{Performance criteria}

To compute the difference in $\Delta (\tau)$, we defined $\tau$ in each dataset as the time at which at least 10\% of the individuals in each group (exposed or unexposed) were still at risk. 
We computed the theoretical values of the $AHR$ and $\Delta(\tau)$ by averaging the estimations obtained, respectively, from univariate Cox PH models ($A$ as the only explanatory covariate) and by equation (\ref{eq_rmst}) where the survival functions were estimated by the Kaplan-Meier estimator, fitted from datasets simulated as above, except $A$ was simulated independently of $L$.\cite{Chatton_2020} We reported the following criteria: (i) the percentage of datasets without convergence ; (ii) the \textcolor{black}{bias either as $\mathrm{E}(\hat \theta)-\theta$ or $100 \times \mathrm{E}(\hat \theta / \theta - 1)$}, where $\theta$ is the estimand of interest; (iii) the mean square error $MSE = \mathrm{E}[(\hat\theta - \theta)^2]$; (iv) the variance estimation bias \textcolor{black}{$VEB = 100 \times (SD(\hat\theta)/\mathrm{E}[\widehat{SD}(\hat\theta)] - 1)$}, where $\mathrm{E}(\widehat{SD}(\bullet))$ is the asymptotic standard deviation and $SD(\bullet)$ is the empirical standard deviation; (v) the empirical coverage rate of the nominal 95\% confidence interval (CI), defined as the percentage of 95\% CIs including $\theta$; (vi) the type I error, defined as the percentage of times the null hypothesis is rejected when the null hypothesis is true; and (vii) the statistical power, defined as the percentage of times the null hypothesis is rejected when the alternative hypothesis is true. \textcolor{black}{We obtained the variances by bootstrap (1000 iterations), as recently recommended by Austin.\cite{Austin_boot}} We computed the Monte Carlo standard errors for each performance measure.\cite{Morris_simulation}

\subsection{Scenarios}

In addition to the two censoring rates and the two effect sizes, we explored three sample sizes: $n=$ 100, 500, and 2000. When the censoring rate was 90\%, we did not investigate the smallest sample size due to the reduced number of events. 

In the main simulations, we considered two sets of covariates: $L=\{L_1, L_2, L_4, L_5\}$ the risk factors of the outcome, or $L=\{L_2, L_5\}$ the true confounders. \textcolor{black}{Therefore, we fitted $g(L)$ as a logistic model and $Q(A,L)$ as a Cox PH model.} 

\textcolor{black}{In a second set of simulations, we aimed to investigate the impact of an omitted confounder on the bias of each method. In addition to the correct set of covariates $L=\{L_1, L_2, L_4, L_5\}$, we defined the incorrect sets of covariates as either $L_{inc}=\{L_1, L_4, L_5\}$ or $\{L_1, L_2, L_4\}$ by respectively omitting $L_2$ or $L_5$, two confounders weakly or strongly associated with both exposure and the outcome. We investigated three scenarios in which the methods used are: (i) $g(L)$ and $Q(A,L_{inc})$, (ii) $g(L_{inc})$ and $Q(A,L)$, and (iii) $g(L_{inc})$ and $Q(A,L_{inc})$.}

\subsection{Software} 

We performed all the analyses using {\tt{R}} version \textcolor{black}{4.0.3}.\cite{R} Source code to reproduce the results is available as Supporting Information on the journal’s web page. To facilitate their use in practice, we have implemented the previous methods in the {\tt{R}} package entitled {\tt{RISCA}} (versions $\geq$ 0.8.1), which is available at {\tt{cran.r-project.org}}.  

\subsection{Results}

\begin{landscape}
\begin{figure}
\centering
\includegraphics[scale=0.45]{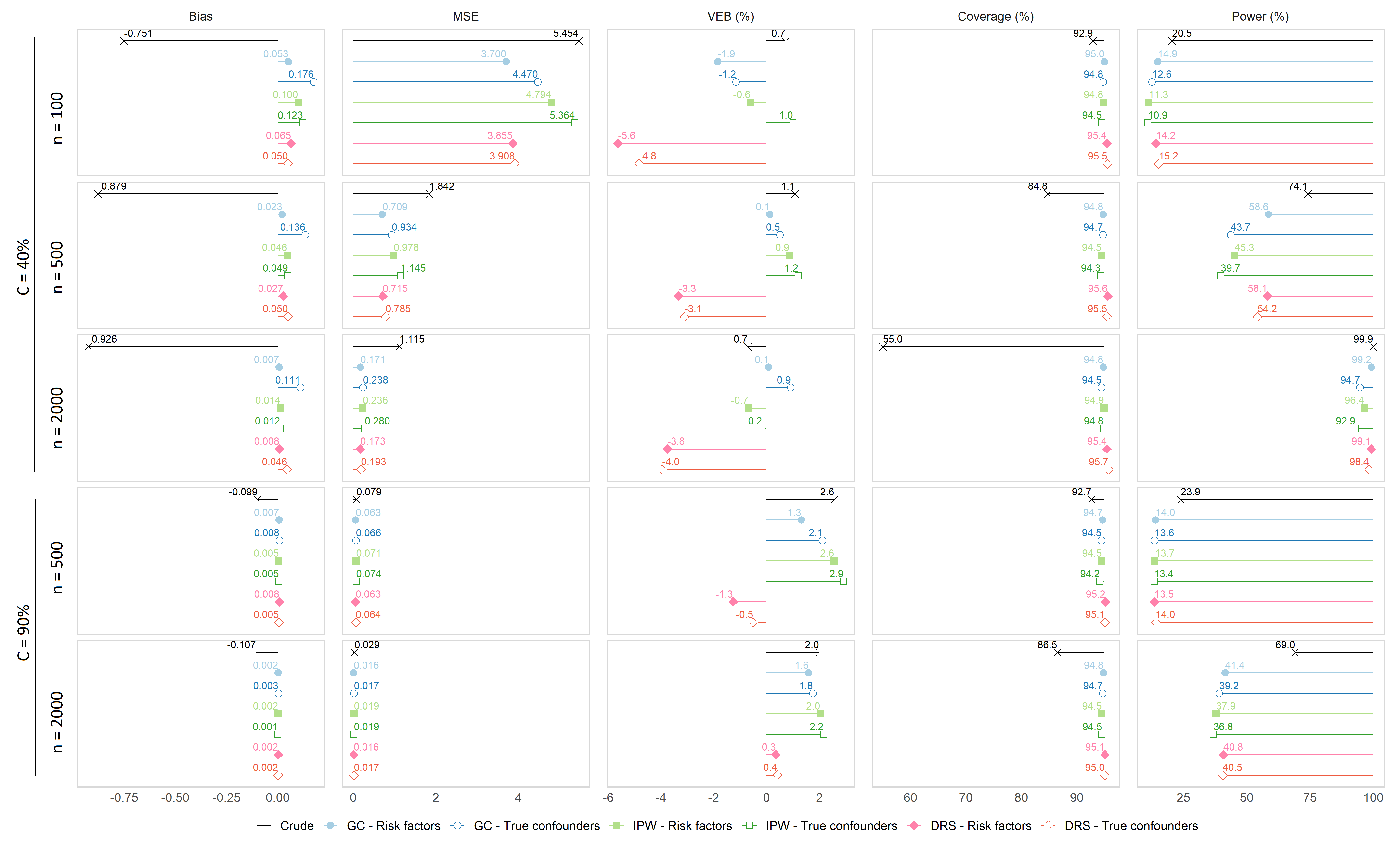}
\caption{Performances of the g-computation (GC), inverse probability weighting (IPW) and Doubly Robust Standardisation (DRS) under the alternative hypothesis to estimate the restricted mean survival times difference at time~$\tau$. $\tau$ equals to 36.6 and 12.9 for censoring rates of 40\% and 90\%, respectively. Theoretical values of restricted mean survival times difference equal to~-1.877 and -0.218 for censoring rates of 40\% and 90\%, respectively. Abbreviations: C, censoring rate; n, sample size.}
\label{resRMST}
\end{figure}

\end{landscape}

The Monte Carlo errors were weak, and we did not encounter any convergence problems. Figure \ref{resRMST} presents the results under the alternative hypotheses for $\Delta$. The results for \textcolor{black}{(i) $\Delta$ under the null hypothesis, and (ii) $AHR$ under the null and alternative hypotheses} were comparable and can be found in the supplementary material available online.

The \textcolor{black}{bias associated with IPW, GC and DRS were similar and close to zero in all scenarios in which we considered all the risk factors or only the true confounders. Nevertheless for GC, the bias of $\Delta$ under the alternative hypothesis was lower considering all the risk factors rather than only the true confounders: 0.007 versus 0.111 for $n=2000$, respectively, but only for a censoring rate of 40\%. In small sample sizes, the bias was higher for IPW than GC and DRS. For instance, when $n=100$ with a censoring rate of 40\% and the risk factors, the bias was 0.100 for GC versus 0.053 and 0.065 for GC and DRS, respectively}.

The GC, when considering all outcome causes, produced the best results in terms of MSE, especially for small sample sizes. For instance, when $n=100$ with a censoring rate of 40\%, the MSE related to the $AHR$ was \textcolor{black}{0.054 for GC versus 0.066 and 0.056 for IPW and DRS, respectively}. When considering only true confounders, these values were \textcolor{black}{0.074, 0.077 and 0.059}, respectively.

Regarding the VEB, the results were slightly better with the GC than IPW, except when $n=100$. \textcolor{black}{DRS underestimated the variance of $\Delta$ under the alternative hypothesis, especially with a censoring rate of 40\%. However, DRS led to a more accurate variance of $\Delta$ with a censoring rate of 90\%. For the $AHR$, DRS led to similar VEBs than GC and IPW.}

All scenarios broadly respected the nominal coverage value of 95\% and the type I error of 5\%. The power was the highest for the GC \textcolor{black}{and DRS}, especially when considering all the risk factors, regardless of the scenario.

\begin{figure}
\centering
\includegraphics[scale=0.45]{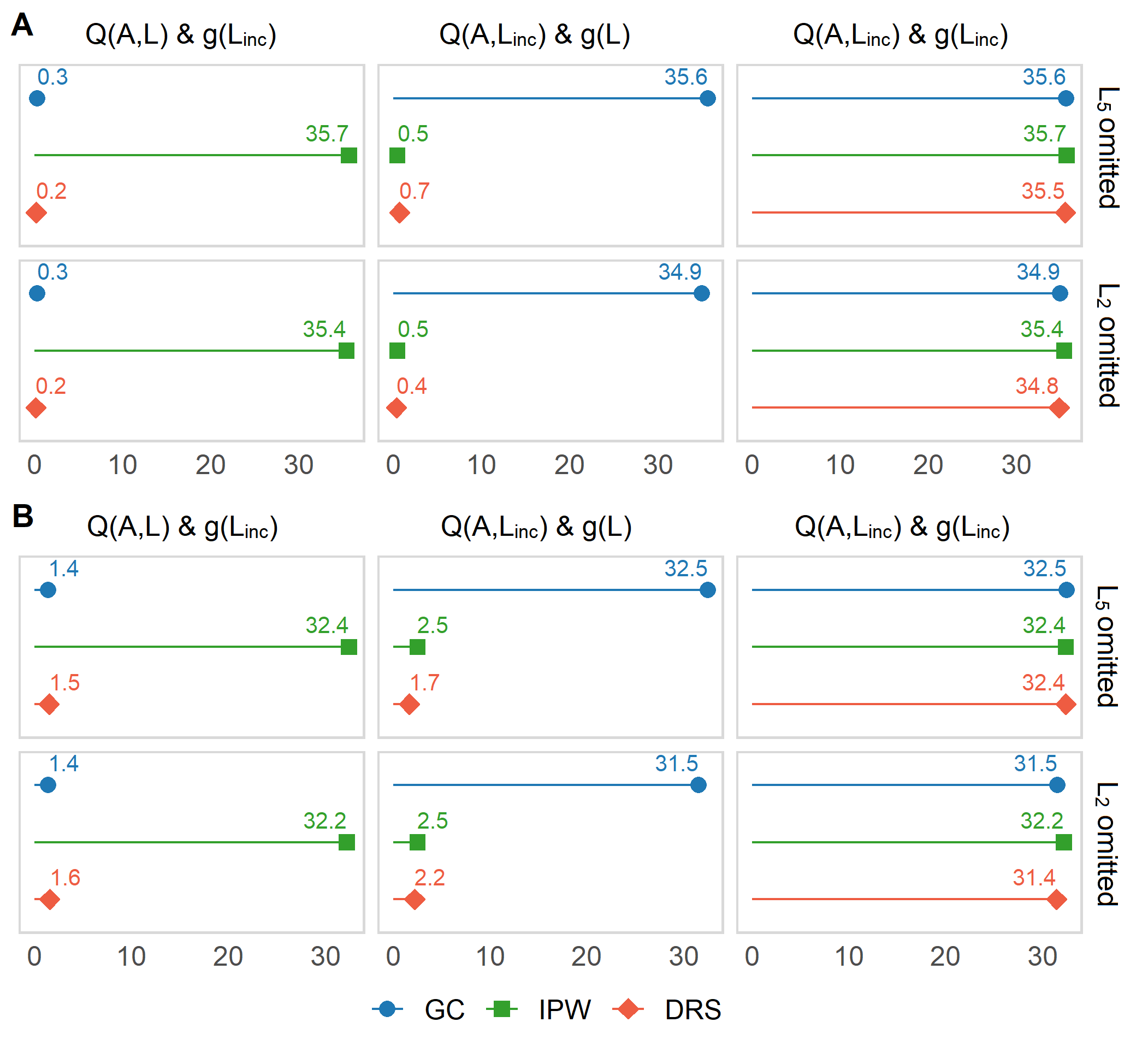}
\caption{Relative bias (\%) of the g-computation (GC), inverse probability weighting (IPW) and Doubly Robust Standardisation (DRS) with an ommitted confounder to estimate: \textbf{A} - the log average hazard ratio; \textbf{B} - the restricted mean survival times difference.}
\label{resU}
\end{figure}

\textcolor{black}{
As expected, omitting a confounder in $g(L)$ or $Q(A,L)$ led to an important bias for IPW and GC, respectively (Figure \ref{resU}). In contrast, DRS remained unbiased when the set of confounders is complete in at least $g(L)$ or $Q(A,L)$. Interestingly, the omission in both $g(L)$ and $Q(A,L)$ did not lead to a higher bias for DRS than GC and IPW. The magnitude of the bias due to the omission of a confounder was similar across methods, ranging from 31.4\% to 35.7\%, for the different estimands and the different strengths of association. Similarly, the sample size and the censoring rate did not significantly change the amplitude of bias (data not shown). 
}

\section{Applications}

We used data from two studies performed for multiple sclerosis and for kidney transplantation.\cite{Laplaud_2019,Masset_2019} We conducted these studies following the French law relative to clinical noninterventional research. Written informed consent was obtained. Moreover, the French commission for data protection approved the collection (CNIL decisions DR-2014-327 and 914184). To guide variable selection, we asked experts which covariates were causes of the exposure or the outcome prognosis to define the causal structure.\cite{VanderWeele_Shpitser_2011} We checked the positivity assumption and the considered covariates balance (see supplementary materials available online). The log-linearity hypothesis of continuous covariates was confirmed in the univariate analysis if the Bayesian information criterion was not reduced using natural spline transformation compared to the inclusion of the covariate in its natural scale. In case of violation, we used a natural spline transformation. We also assessed the PH assumption via the Grambsch-Therneau test at a significance level of 5\%. For simplicity, we performed complete case analyses.

\subsection{Dimethylfumarate versus Teriflunomide to prevent relapse in multiple sclerosis}

With the increasing number of available drugs for preventing relapses in multiple sclerosis and the lack of head-to-head randomised clinical trials, Laplaud \textit{et al.}\cite{Laplaud_2019} aimed to compare Teriflunomide (TRF) and Dimethylfumarate (DMF) using data from the multicentric cohort OFSEP. We reanalysed the primary outcome, defined as the time-to-first relapse. We presented the cohort characteristics of 1770 included patients in Table 1: 1057 patients were in the DMF group (59.7\%) versus 713 in the TRF group (40.3\%). Approximately 39\% of patients (40\% in the DMF group versus 38\% in the TRF group) had at least one relapse during follow-up.

\begin{landscape}
\begin{figure}
\centering
\includegraphics[scale=0.75]{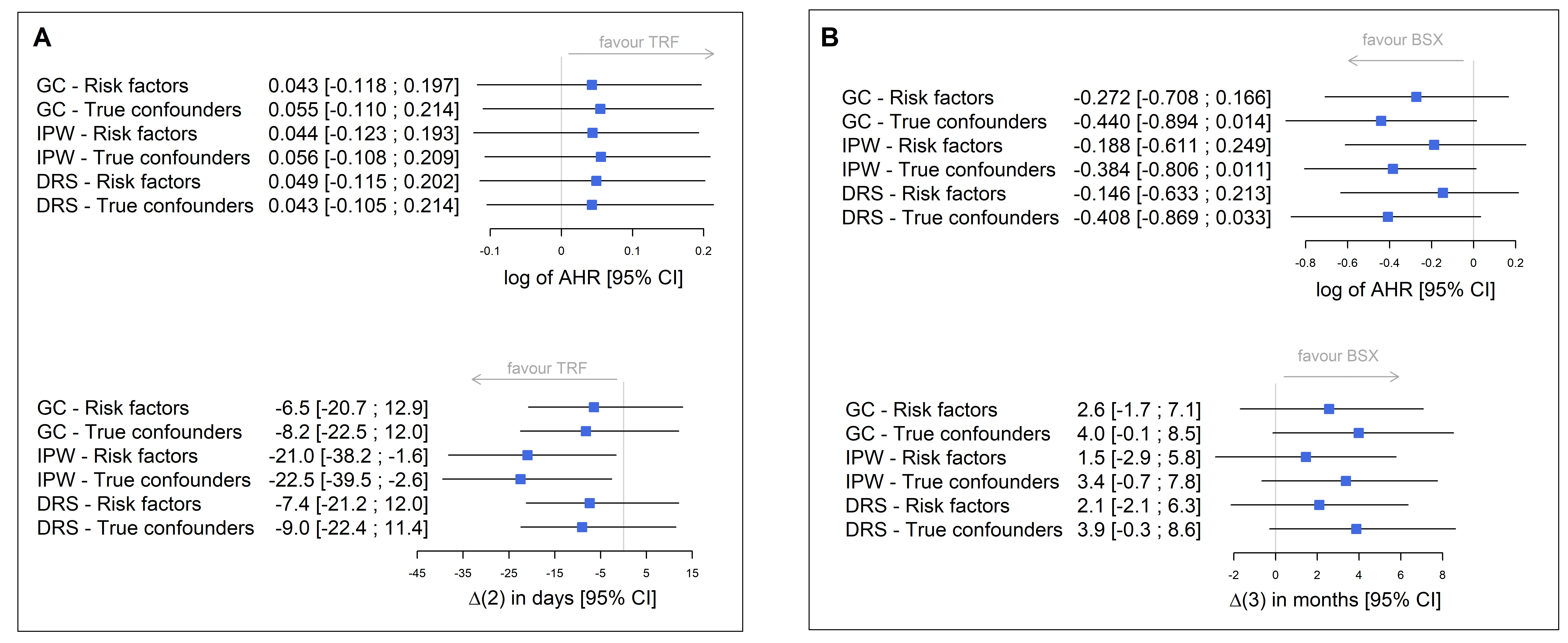}
\caption{Comparison of: \textbf{A} - Dimethylfumarate and Teriflunomide (TRF) for the time-to-first relapse of multiple sclerosis; \textbf{B} - Basiliximab (BSX) and Thymoglobulin for the occurrence of a cardiovascular complication after a kidney's transplantation.}
\label{forCVC}
\end{figure}
\end{landscape}

We presented the confounders-adjusted results in the left panel of Figure \ref{forCVC}. \textcolor{black}{The different set of covariates did not significantly change the results}. For the difference in RMST, the width of the 95\% CI was larger for the IPW. For instance, when we considered all the risk factors, the CI of IPW had a width of \textcolor{black}{36.6 days versus 33.6 days and 33.2 days for GC and DRS, respectively}.

The conclusion of no significant difference between TRF and DMF was unaffected by the method \textcolor{black}{for the $AHR$. In contrast, the IPW led to a protective effect of the TRF compared to DMF in terms of $\Delta$ at two years, while the GC and the DRS remained consistent with the $AHR$ conclusions. By taking into account the risk factors, the use of IPW concluded to a gain of 21 days without relapse at two years versus 6.5 days and 7.4 days for GC and DRS, respectively. Owing to the similar performances of the three methods under the null hypothesis and because unmeasured confounding cannot be an issue here, we suppose that the difference can be explained by a misspecification of $g(L)$.}

\subsection{Basiliximab versus Thymoglobulin to prevent post-transplant complications}

Amongst non-immunised kidney transplant recipients, one can expect similar rejection risk between Thymoglobulin (ATG) and Basiliximab (BSX), two possible immunosuppressive drugs proposed as induction therapy. However, ATG may be associated with higher serious events, especially in the elderly. We aimed to reassess the difference in cardiovascular complications in ATG versus BSX patients.\cite{Masset_2019} Table 2 describes the 383 included patients from the multicentric DIVAT cohort: 204 patients were in the BSX group (53.3\%) versus 179 in the ATG group (46.7\%). Approximately 30\% of patients (29\% in the BSX group and 31\% in the ATG group) had a least one cardiovascular complication during follow-up. The median follow-up time was 1.8 years (min: 0.0; max: 8.2).

In the right panel of Figure \ref{forCVC}, we presented the confounders-adjusted RMST differences for a cohort followed up to three years. \textcolor{black}{The results obtained were similarly sensitive to the considered set of covariates. Indeed, the upper (respectively lower) bound of the 95\% CIs of the $AHR$ (respectively $\Delta$) was closer to zero when only the risk factors were considered. Nevertheless, the conclusion remained identical whatever the method used and the estimand targeted: we were unable to conclude to a difference in cardiovascular complications between the ATG and BSX patients.}

\section{Discussion}

We aimed to explain and compare the performances of the GC\textcolor{black}{, IPW and DRS} for estimating causal effects in time-to-event analyses with time-invariant confounders. We focused on the average HR and the RMST difference. The results of the simulations showed that the \textcolor{black}{three} methods performed similarly in terms of \textcolor{black}{bias}, VEB, coverage rate, and type I error rate. Nevertheless, \textcolor{black}{both the GC and the DRS} outperformed the IPW in terms of statistical power, even when the censoring rate was high. Furthermore, the simulations showed that \textcolor{black}{$Q(A,L)$} should preferentially include all the risk factors to ensure a smaller bias \textcolor{black}{due to the self-induced selection}. The main advantage of using the GC is the gain in statistical power. \textcolor{black}{DRS is also an interesting method due to its double robustness property at the cost of a small loss of efficiency.}

\textcolor{black}{We have overcome the self-induced selection in HR \textit{i.e.}, the non-collapsibility due to the presence of a cause of the time-to-event independent of the exposure, by averaging the time-dependant HR. The $AHR$ is furthermore a valid causal estimand because it creates a contrast between a function of the potential outcomes under the two exposures $A=1$ or $A=0$.\cite{Martinussen_2020} 
Surprisingly, we reported a higher bias of the RMST difference by GC when we only considered the true confounders. The difference in survival, such as RMST, is only collapsible in a fully continuous-time context, \textit{i.e.}, with infinitely short time intervals.\cite{Sjolander_2016} For the GC, we computed the survival probability for each observed event time leading to small intervals. Therefore, we could observe slight, but still present, bias due to non-collapsibility when the risk factors were not considered. We did not observe such bias with the higher censoring rate due to the scarcity of events.\cite{Aalen_2015}} 

\textcolor{black}{While the first application to multiple sclerosis highlighted differences between the duo GC/DRS and the IPW, the second one in kidney transplantation illustrated the importance of the set of covariates to consider. With all risk factors, we concluded with more confidence that there was not a significant difference between Basiliximab and Thymoglobulin. In contrast, when using only the true confounders, the bound of the 95\% CIs were close to zero. This again highlights} the fact that even if a risk factor is balanced between the exposure groups at baseline, it could become unbalanced over time (as illustrated in Figure~\ref{Fig1}, panel B).

Nevertheless, the higher power of the GC is counterbalanced \textcolor{black}{by three points. First, IPW allows us to easily check positivity near-violations by plotting the individual weights.\cite{Petersen_2012} Second, although the lack of positivity directly affects the estimation of $g(L)$, the GC is also impacted by the resulting lack of support to properly estimate $Q(A,L)$ which can be qualified as an extrapolation issue. An extension of our work is to explore the robustness of the previous methods in the presence of such near-violations.} 
Third, the need for bootstrapping to estimate its variance, analytic estimators that are available for the IPW.\cite{Conner_Trinquart_2019, Hajage_2018} In practice, we must emphasise that bootstrapping the entire estimation procedure has the advantage of valid post-selection inference.\cite{Efron_2014} Furthermore, data-driven methods for variables selection, such as the super learner, have recently been developed and may represent a promising perspective when full clinical knowledge is unavailable.\cite{Blakely_2019, LeBorgne_2021} With such a data-driven covariates selection, the use of GC also represents a partial solution to prevent the selection of instrumental variables since it is independent of the exposure modelling\textcolor{black}{, but DRS can avoid potential residual confounding due to the omission of a confounder weakly associated with the outcome in $Q(A,L)$.\cite{Kreif_DiazOrdaz_2019} While DREs have been criticised because they can amplify the bias when the two working models are misspecified,\cite{Kang_Schafer_2007, Tan_2007} DRS and TMLEs do not cause such bias amplification.\cite{Vansteelandt_Keiding_2011, Kreif_2016} Some TMLEs have been proposed to estimate either the RMST difference or survival functions.\cite{Diaz_2019, Benkeser_2018} Unfortunately, they require the discretisation of the times exacerbating the non-collapsibility of the estimands.\cite{Sjolander_2016}}

The methods studied here are not the only available methods to estimate the causal effect. For instance, Conner \textit{et al.}\cite{Conner_Trinquart_2019} compared the performances of IPW with that of other regression-based methods. Overall, the statistical performances were similar. However, the advantage of \textcolor{black}{the studied methods} compared to other ones is the visualisation of the confounder-adjusted results in terms of the survival curve or an indicator such as RMST.

Our study has several limitations. First, the results of the simulations and applications are not a theoretical argument for generalising to all situations. Second, we studied only logistic and Cox PH regression: other \textcolor{black}{working} models could be applied. Keil and Edwards\cite{Keil_Edwards_2018} proposed a review of possible \textcolor{black}{models for $Q(A,L)$} with a time-to-event outcome. Third, we considered only a reduced number of covariates, which could explain the abovementioned equivalence between the GC and the IPW with the extreme censoring rate. Last, we did not consider competing events or time-varying confounders that require specific estimation methods.\cite{Young_2020, Daniel_2013}

To conclude, by means simulation and two applications on real
datasets, this study tended to show the \textcolor{black}{lower} power of the
\textcolor{black}{IPW} compared to \textcolor{black}{GC and DRS} to estimate the causal effect with time-to-event outcomes. All the risk factors should be considered in \textcolor{black}{GC to overcome the self-induce selection bias}. Our work is a continuation of the emerging literature that questions the near-exclusivity of PS-based methods in causal inference.

\section*{Acknowledgements}
The authors would like to thank the members of DIVAT and OFSEP Groups for their involvement in the study, the physicians who helped recruit patients, the clinical research associates who participated in the data collection and all patients who participated in this study. We also thank David Laplaud and Magali Giral for their clinical expertise as well as Gabriel Danelian for language corrections. The analysis and interpretation of these data are the responsibility of the authors. This work was partially supported by a public grant overseen by the French National Research Agency (ANR) to create the Common Laboratory RISCA (Research in Informatic and Statistic for Cohort Analyses, {\tt{www.labcom-risca.com}}, reference: ANR-16-LCV1-0003-01) involving the development of Plug-Stat software. 

\section*{Declaration of conflicting interests}
The authors declared no potential conflicts of interest with respect to the research, authorship, and/or publication of this article.

\section*{Funding}
Arthur Chatton obtained a grant from IDBC for this work.
Other authors received no financial support for the research, authorship, and/or publication of this article.

\bibliographystyle{SageV}

\newpage
%Tables

\begin{table}
\small\sf\centering 
\caption{Description of the multiple sclerosis cohort according to the treatment group.} 
\begin{tabular}{lccccccc}
\toprule
& \multicolumn{2}{c}{Overall (n=1770)}   & \multicolumn{2}{c}{TRF (n=713)}  & \multicolumn{2}{c}{DMF (n=1057)} & p-value \\ 
     & \textit{n}           & \textit{\%}      & \textit{n}    & \textit{\%}   & \textit{n}    & \textit{\%}   &        \\ 
\midrule
Male recipient  & 485     & 27.4 & 202  & 28.3 & 283  & 26.8 & 0.4713            \\
Disease modifying therapy before initiation  & 1004\textcolor{white}{-}    & 56.7 & 395  & 55.4 & 609  & 57.6 & 0.3560            \\
Including Interferon                                                &         &      & 237  &      & 369  &      &                   \\
\textcolor{white}{Including} Glatiramer Acetate                                                  &         &      & 158  &      & 240  &      &                   \\
Relapse within the year before initiation & 981     & 55.4 & 346  & 48.5 & 635  & 60.1 & $<$0.0001\textcolor{white}{$<$} \\
Relapse within the two years before initiation & 1227\textcolor{white}{-}    & 69.3 & 444  & 62.3 & 783  & 74.1 & $<$0.0001\textcolor{white}{$<$} \\
Gado. Positive lesion on MRI at baseline  & 601     & 34.0 & 207  & 29.0 & 394  & 37.3 & 0.0003            \\
Center with more than 50 included patients  & 1612\textcolor{white}{-}    & 91.1 & 653  & 91.6 & 959  & 90.7 & 0.5354            \\
At least one relapse at two-year post-initiation & 527     & 29.8 & 200  & 28.1 & 327  & 30.9 & 0.1928            \\
\\
     & \textit{mean}           & \textit{sd}      & \textit{mean}    & \textit{sd}   & \textit{mean}    & \textit{sd}   &        \\ 
Patient age at multiple sclerosis onset (years) & 31.7    & \textcolor{white}{-}9.7  & 32.9 & \textcolor{white}{-}9.8  & 30.9 & \textcolor{white}{-}9.5  & $<$0.0001\textcolor{white}{$<$} \\
Patient age at initiation (years)   & 39.3    & 10.7 & 41.3 & 10.8 & 38.0 & 10.5 & $<$0.0001\textcolor{white}{$<$} \\
Disease duration (years)  & \textcolor{white}{-}7.6     & \textcolor{white}{-}7.4  & \textcolor{white}{-}8.4  & \textcolor{white}{-}7.8  & \textcolor{white}{-}7.1  & \textcolor{white}{-}7.0  & 0.0003            \\
EDSS level at initiation   & \textcolor{white}{-}1.7     & \textcolor{white}{-}1.3  & \textcolor{white}{-}1.7  & \textcolor{white}{-}1.3  & \textcolor{white}{-}1.7  & \textcolor{white}{-}1.2  & 0.9885            \\
Number of relapses in the previous year   & \textcolor{white}{-}0.7    & \textcolor{white}{-}0.8 & \textcolor{white}{-}0.6 & \textcolor{white}{-}0.7 & \textcolor{white}{-}0.8 & \textcolor{white}{-}0.8 & $<$0.0001\textcolor{white}{$<$} \\
Number of relapses in the two previous years   & \textcolor{white}{-}1.0    & \textcolor{white}{-}1.0 & \textcolor{white}{-}0.9 & \textcolor{white}{-}0.9 & \textcolor{white}{-}1.1 & \textcolor{white}{-}1.0 & $<$0.0001\textcolor{white}{$<$} \\
\bottomrule
\multicolumn{8}{p{15cm}}{No variable have missing data. \newline
Abbreviations: DMF, Dimethylfumarate; EDSS, Expanded Disability Status Scale; Gado, Gadolinium; MRI, Magnetic resonance imaging; MS, Multiple sclerosis; sd, Standard deviation; and TRF, Teriflunomide.}
\end{tabular}
\label{tab1:sep}
\end{table}

\begin{table}
\caption{Description of the kidney's transplantation cohort according to the induction therapy.} 
\small\sf\centering 
\begin{tabular}{lcccccccccc}
\toprule
 & \multicolumn{3}{c}{Overall (n=383)}                                & \multicolumn{3}{c}{ATG (n=179)} & \multicolumn{3}{c}{BSX (n=204)} & p-value   \\ 
     & \textit{missing}          & \textit{n}           & \textit{\%}      & \textit{missing} & \textit{n}    & \textit{\%}   & \textit{missing} & \textit{n}    & \textit{\%}   &        \\ \midrule
Male recipient                               & 0           & 284         & 74.2    & 0  & 137  & 76.5 & 0  & 147  & 72.1 & 0.3180 \\
Recurrent causal nephropathy                 & 0           & \textcolor{white}{-}63          & 16.4    & 0  & \textcolor{white}{-}29   & 16.2 & 0  & \textcolor{white}{-}34   & 16.7 & 0.9024 \\
Preemptive transplantation                   & 1           & \textcolor{white}{-}61          & 16.0    & 1  & \textcolor{white}{-}18   & 10.1 & 0  & \textcolor{white}{-}43   & 21.1 & 0.0035 \\
History of diabetes                          & 0           & 123         & 32.1    & 0  & \textcolor{white}{-}64   & 35.8 & 0  & \textcolor{white}{-}59   & 28.9 & 0.1530 \\
History of hypertension                      & 0           & 327         & 85.4    & 0  & 150  & 83.8 & 0  & 177  & 86.8 & 0.4124 \\
History of vascular disease                  & 0           & 109         & 28.5    & 0  & \textcolor{white}{-}53   & 29.6 & 0  & \textcolor{white}{-}56   & 27.5 & 0.6405 \\
History of cardiac disease                   & 0           & 153         & 39.9    & 0  & \textcolor{white}{-}75   & 41.9 & 0  & \textcolor{white}{-}78   & 38.2 & 0.4651 \\
History of cardiovascular disease            & 0           & 203         & 53.0    & 0  & \textcolor{white}{-}99   & 55.3 & 0  & 104  & 51.0 & 0.3973 \\
History of malignancy                        & 0           & \textcolor{white}{-}94          & 24.5    & 0  & \textcolor{white}{-}42   & 23.5 & 0  & \textcolor{white}{-}52   & 25.5 & 0.6457 \\
History of dyslipidemia                      & 0           & 220         & 57.4    & 0  & \textcolor{white}{-}92   & 51.4 & 0  & 128  & 62.7 & 0.0250 \\
Positive recipient CMV serology              & 5           & 230         & 60.8    & 4  & 119  & 68.0 & 1  & 111  & 54.7 & 0.0082 \\
Male donor                                   & 0           & 187         & 48.8    & 0  & \textcolor{white}{-}93   & 52.0 & 0  & \textcolor{white}{-}94   & 46.1 & 0.2510 \\
ECD donor                                    & 1           & 372         & 97.4    & 1  & 172  & 96.6 & 0  & 200  & 98.0 & 0.5244 \\
Use of machine perfusion                     & 12\textcolor{white}{-}          & 208         & 54.3    & 6  & \textcolor{white}{-}86   & 48.0 & 6  & 122  & 59.8 & 0.0684 \\
Vascular cause of donor death                & 0           & 275         & 71.8    & 0  & 126  & 70.4 & 0  & 149  & 73.0 & 0.5655 \\
Donor hypertension                           & 11\textcolor{white}{-}          & 224         & 60.2    & 9  & 103  & 60.6 & 2  & 121  & 59.9 & 0.8927 \\
Positive donor CMV serology                  & 0           & 240         & 62.7    & 0  & 115  & 64.2 & 0  & 125  & 61.3 & 0.5486 \\
Positive donor EBV serology                  & 1           & 370         & 96.9    & 1  & 172  & 96.6 & 0  & 198  & 97.1 & 0.8102 \\
HLA-A-B-DR incompatibilities \textgreater{}4 & 5           & \textcolor{white}{-}97          & 25.7    & 3  & \textcolor{white}{-}41   & 23.3 & 2  & \textcolor{white}{-}56   & 27.7 & 0.3256 \\
\\
  &   & \textit{mean}           & \textit{sd}      &  & \textit{mean}    & \textit{sd}   &  & \textit{mean}    & \textit{sd}   &        \\
Recipient age (years)                        & 0           & 70.8        & \textcolor{white}{-}4.8     & 0  & 70.5 & \textcolor{white}{-}4.8  & 0  & 71.0 & \textcolor{white}{-}4.8  & 0.3733 \\
Recipient BMI (kg/m$^2$)                       & 3           & 26.7        & \textcolor{white}{-}4.0     & 3  & 26.9 & \textcolor{white}{-}4.2  & 0  & 26.5 & \textcolor{white}{-}3.9  & 0.2796 \\
Duration on waiting list (months)            & 16\textcolor{white}{-}          & 16.5        & 19.0    & 11\textcolor{white}{-} & 17.9 & 18.9 & 5  & 15.4 & 19.1 & 0.2082 \\
Donor age (years)                            & 1           & 72.7        & \textcolor{white}{-}8.8     & 1  & 72.1 & 10.0 & 0  & 73.1 & \textcolor{white}{-}7.5  & 0.2739 \\
Donor creatininemia ($\mu$mol/L)                 & 1           & 82.9        & 39.5    & 0  & 85.5 & 41.0 & 1  & 80.7 & 38.0 & 0.2331 \\
Cold ischemia time (hours)                   & 3           & 15.6        & \textcolor{white}{-}5.0     & 1  & 15.9 & \textcolor{white}{-}5.2  & 2  & 15.3 & \textcolor{white}{-}4.8  & 0.2820\\ 
\bottomrule
\multicolumn{11}{p{17cm}}{Abbreviations: ATG, Thymoglobulin; BMI, Body mass index; BSX, Basiliximab; CMV, Cytomegalovirus; EBV, Epstein-Barr virus; ECD, Expanded criteria donor; HLA, Human leucocyte antigen; and sd, Standard deviation.}
\end{tabular}
\label{tab2:cvc}
\end{table}

\newpage
%Figures

\newpage

\newpage

\end{document}